# A Simple Distortion Calibration method for Wide-Angle Lenses Based on Fringe-pattern Phase Analysis

WEISHUAI ZHOU,[1,†] JIAWEN WENG,[2,†] JUNZHENG PENG,[1] JINGANG ZHONG[1,*]

[1] *Department of Optoelectronic Engineering, Jinan University, Guangzhou 510632, China*
[2] *Department of Physics, South China Agricultural University, Guangzhou 510642, China*
*†The authors contributed equally to this work.*
*\*Corresponding author: tzjg@jnu.edu.cn*

A distortion calibration method for wide-angle lens is proposed based on fringe-pattern phase analysis. Firstly, according to the experimental result of the radial distortion of the image not related to the recording depth of field, but depending on the field of view angle of the wide-angle lens imaging system, two-dimensional image distortion calibration is need to be considered. Four standard sinusoidal fringe-patterns with phase shift step of $\pi/2$, which are used as calibration templates, are shown on a Liquid Crystal Display screen, and captured by the wide-angle lens imaging system. A four-step phase-shifting method is employed to obtain the radial phase distribution of the distorted fringe-pattern. Wavelet analysis is applied for the analysis of the instantaneous frequency to show the fundamental frequency of the fringe-pattern in the central region being unchanged. Performing numerical calculation by the central 9 points of the central row of the fringe-pattern, we can get the undistorted radial phase distribution, so, the radial modulated phase is computed. Finally, the radial distortion distribution is determined according to the radial modulated phase. By employing a bilinear interpolation algorithm, the wide-angle lens image calibration is achieved. There is no need to establish any kind of image distortion model for the proposed method. There is no projecting system in the experimental apparatus, which avoids projection shadow problems, and no need to align with the center of the template for distortion measurement for the proposed method. Theoretical description, numerical simulation and experimental results show that the proposed method is simple, automatic and effective.

## 1. INTRODUCTION

Wide-angle lens [1] is widely utilized in various applications, such as computer vision, sky observation [2], military scenario [3], medical imaging system [4], and so on. However, wide-angle lens distortion bends straight lines into circular arcs. It would reduce the accuracy of quantitative measurements in determining the geometric position and size of objects in the image, and even cause misjudgment in the process of image recognition. Therefore, it is especially crucial to preform distortion correction.

Brown [5] proposed the radial, decentering and prism distortion model for image distortion. The radial distortion plays a dominant role in wide-angle lens. The radial distortion, including barrel distortion and pincushion distortion according to different focal lengths, is classified into polynomial model [5] and division model [6]. Most conventional distortion calibration algorithms use measuring template methods, which includes two methods. One is plumb-line method, relying on a 2D single template, [7-9]. Plumb-line method call for a template needing sufficient number of straight lines. Another is multiple view calibration [6, 10-12] method, utilizing the pin-hole and lens distortion models together, which requires a set of images. Both of the above methods need to establish distortion models, extract many feature points or lines to determine distortion parameters, for image calibration.

Here, a simple and automatic calibration method for wide-angle lens based on fringe-pattern phase analysis is proposed, which does not need to establish distortion model and extract feature points or lines either, and can calculate the distortion value of each pixel of the distorted image, correcting the image quickly and accurately. According to the experimental result of the radial distortion of the image not related to the recording depth of field, but depending on the field of view angle of the wide-angle lens imaging system, a simple and effective two-dimensional (2-D), rather than the three-dimensional (3-D), image distortion calibration is need to be considered. Four one-dimensional (1-D) sinusoidal fringe-patterns with phase shift step of $\pi/2$, which are shown on the Liquid Crystal Display (LCD) screen, are employed as calibration templates, and captured by the wide-angle lens imaging system. A four-step phase-shifting method is used for the phase de-modulation to obtain the radial phase distribution of the distorted fringe-pattern. And then, wavelet analysis shows that the instantaneous frequency of the fringe-pattern in the central region remains unchanged. So, by performing numerical calculation according to the central 9 points of the fringe-pattern, we can get the undistorted radial phase distribution. Therefore, the radial modulated phase can be computed by the difference between the distorted and undistorted radial phase distribution. Finally, the radial distortion distribution is determined. By employing a bilinear interpolation algorithm, the wide-angle lens image calibration is achieved.

There is no projecting system in the experimental apparatus, which avoids projection shadow problems well. By introducing one-dimensional (1-D) carrier-fringes as the calibration templates, there is no need to align with the center of the template for distortion measurement. Compared with Fourier analysis method, the four-step phase-shifting method with the best spatial localization merit can provide the high precision phase distribution of the distorted fringe-pattern. That is the most important step for the measurement of the image distortion. Section 2 and 3 explain the related works of the proposed method, including the principle of distortion calibration and the measurement of the radial distortion. Section 4 shows the numerical calculation in detail and the experimental results. Lastly, conclusions showing the advantages of the proposed method are given in Section 5.

## 2. DISTORTION CALIBRATION PRINCIPLE

In this paper, a simple and effective 2-D image distortion calibration method, is proposed according to the experimental result of the radial distortion of the image not related to the recording depth of field, but depending on the field of view angle of the wide-angle lens imaging system.

Assuming that the undistorted image is $R(x, y)$, the distorted image is $D(x', y')$. $(x, y)$ and $(x', y')$ are the corresponding coordinates respectively. The relationship between these two coordinates due to distortion can be expressed as

$$\begin{cases} x = o_1(x', y') \\ y = o_2(x', y') \end{cases}. \quad (1)$$

For the wide-angle lens imaging system we discuss here, the radial distortion is assumed to be circular symmetric about the center of the image when the wide-angle lens is perpendicular to the calibration template. In addition, the following experiments in Section 4 show that the radial distortion of the image is not related to the recording depth of field, but depends on the field of view angle of the wide-angle lens imaging system, i.e., the distance of the pixel point from the center in the image plane. So, we could consider a 2-D, rather than 3-D, image distortion calibration. It means that we just use the distance from the target point to the central point $(x_0, y_0)$ of the image to calculate the distortion value for the whole image. The relationship between the distorted image coordinate and undistorted image coordinate can be expressed simply as

$$r = o(r'). \quad (2)$$

Where $r(x, y) = \sqrt{(x-x_0)^2 + (y-y_0)^2}$ and $r'(x', y') = \sqrt{(x'-x_0)^2 + (y'-y_0)^2}$ represent the distance of the point $(x, y)$ and $(x', y')$ from the central point of the image respectively. If the mapping relationship between the distorted image coordinates and the undistorted image coordinates is established, the image could achieve calibration.

Suppose that $C[r(x, y)]$ is the gray value of a pixel point at position $(x, y)$ in the calibrated image. It corresponds to the pixel point $(x', y')$ at position with the same gray value $D[r'(x', y')]$ in the distorted image. So, there is

$$C[r(x, y)] = D[r'(x', y')] = D[r(x, y) + \Delta r]. \quad (3)$$

Where $\Delta r$ is the corresponding radial distortion value. Since the radial distortion is rotationally symmetric with respect to the center of the distorted image, what we need to do is to determine the radial distortion distribution along the positive direction from the central point of the image for image calibration.

It should be noticed that the coordinate value $(x, y)$ of a pixel point of the calibrated image would be an integer, but the coordinate value $(x', y')$ calculated by Eq. (3) might not be an integer. Therefore, it is necessary to perform a gray interpolation algorithm. A bilinear interpolation algorithm is employed here not only because of the simple and convenient calculation process, but also due to its ability of overcoming the problem of grayscale discontinuity. Suppose that $x_1$ and $y_1$ are the nearest integers less than or equal to $x'$ and $y'$ respectively, so, $(x', y')$ falls within the two-dimensional region controlled by four pixels of $(x_1, y_1)$, $(x_1+1, y_1)$, $(x_1, y_1+1)$ and $(x_1+1, y_1+1)$. The calculation formula of the bilinear interpolation algorithm is

$$\begin{aligned} C[r(x, y)] = &(1-\alpha)(1-\beta)D[r(x_1, y_1)] + \alpha(1-\beta)D[r(x_1+1, y_1)] \\ &+ (1-\alpha)\beta D[r(x_1, y_1+1)] + \alpha\beta D[r(x_1+1, y_1+1)]. \end{aligned} \quad (4)$$

Where $\alpha = x' - x_1$ and $\beta = y' - y_1$.

## 3. RADIAL DISTORTION MEASUREMENT

### A. Fringe-pattern phase analysis

A fringe-pattern along the longitudinal direction is employed as calibration template, as shown in Fig. 1. The dotted lines and the solid lines are corresponding to the carrier-fringe stripes before and after distortion respectively. In this paper, four sinusoidal fringe-patterns with phase shift of $0, \pi/2, \pi, 3\pi/2$ are employed. The intensity distributions of the fringe-patterns are

$$I_1(x, y) = A(x, y) + B(x, y)\cos[\varphi(x, y)], \quad (5)$$

$$I_2(x, y) = A(x, y) + B(x, y)\cos[\varphi(x, y) + \pi/2], \quad (6)$$

$$I_3(x, y) = A(x, y) + B(x, y)\cos[\varphi(x, y) + \pi], \quad (7)$$

$$I_4(x, y) = A(x, y) + B(x, y)\cos[\varphi(x, y) + 3\pi/2]. \quad (8)$$

Where $A(x, y)$ is the background intensity, $B(x, y)/A(x, y)$ is the contrast of the fringe-pattern; $\varphi(x, y)$ is the phase of fringe-pattern, which contains the distortion information of the image. By employing a four-step phase-shifting method for phase de-modulation of the central rows of the four fringe-patterns, the radial wrapped phase distribution can be acquired from the four distorted fringe-patterns as

$$\varphi(x, y) = \arctan\left[\frac{I_4(x, y) - I_2(x, y)}{I_1(x, y) - I_3(x, y)}\right]. \quad \textbf{(9)}$$

By employing the unwrapping algorithm, the radial phase distribution of the distorted fringe-pattern could be obtained.

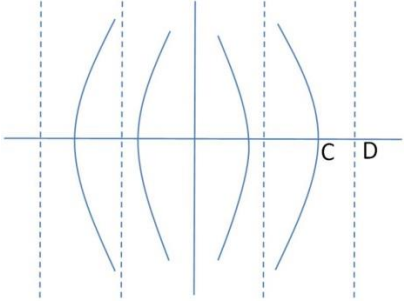

Fig. 1. Distortion schematic diagram.

**B. Radial distortion measurement**

The instantaneous frequency of the fringe-pattern in the central region remains unchanged shows that there is little image distortion in this region. The following wavelet analysis for the frequency analysis of the fringe-pattern shows the result in Section 4. So, we can perform numerical linear fitting according to the central 9 points of radial distorted phase, which is calculated by Eq. (9), to get the radial phase distribution of the undistorted fringe-pattern as

$$\varphi_{undistorted}(x) = kx + \varphi_0. \quad \textbf{(10)}$$

Where $k = 2\pi f_0$ is the linear fitting parameter, $f_0$ is the fundamental frequency, $\varphi_0$ is a constant parameter. The radial modulated distribution can be obtained as

$$\Delta\varphi(x) = \varphi_c(x) - \varphi_{undistorted}(x). \quad \textbf{(11)}$$

Where $\varphi_c(x)$ is the radial phase of the distorted fringe-pattern calculated by the four-step phase-shifting method. According to the geometry shown in Fig. 1, the relationship of the radial distortion value $\Delta r$ and the corresponding radial modulated phase $\Delta\varphi$ can be expressed as

$$\Delta r = \overline{CD} = \frac{\Delta\varphi}{2\pi f_0}. \quad \textbf{(12)}$$

So, by employing the fringe-pattern phase analysis, the radial distortion distribution can be determined. Finally, the image calibration is achieved according to Eq. (3) and Eq. (4).

## 4. EXPERIMENTAL RESULT

The experiment is performed employing the apparatus depicted in Fig. 2, where the optical axis is perpendicular to the calibration templates shown on LCD. There is no projecting system, so, it can avoid projection shadow problems well. Four sinusoidal fringe-patterns with $\pi/2$ phase shift shown on LCD (FunTV D49Y) screen are captured by a wide-angle lens (Theia MY125M, FOV 137°, depth of field: 10cm to infinity) that will be calibrated. The captured images shown in Fig. 3, are employed for numerical analysis in the personal computer(PC). The size of the fringe-patterns is 1080×1920 pixels. The analysis process can be described as the following steps:

Step1: Employ a four-step phase-shifting method for phase de-modulation of the central rows of the four fringe-patterns and perform the unwrapping algorithm to obtain the radial phase distribution of the distorted fringe-pattern. Cubic polynomial fitting is performed to smooth the radial distorted phase distribution line to achieve de-noising. Figure 4 shows the gray value distribution of the central row of the second fringe-pattern shown in Fig.3. We can find that the intensity in the central region is much higher than that at both sides. When employing fourier analysis, great error would be introduced into the phase analysis. The four-step phase-shifting method due to its best spatial localization merit can provide the high precision phase distribution of the distorted fringe-pattern. That is the most important step for the measurement of the image distortion.

Step2: Numerical calculate the radial phase distribution of the undistorted fringe-pattern. Wavelet analysis for the instantaneous frequency analysis of the fringe-pattern is performed to prove the fringe pattern in the central region undistorted roughly. By detecting the maximum value of the wavelet coefficient of the central row of the fringe-pattern at each position, the corresponding instantaneous frequency can be obtained as shown in Fig. 5. Figure 5 shows that the instantaneous frequency of the fringe-pattern in the central region remains unchanged. So there is little image distortion in this region. Numerical linear fitting is performed according to the central 9 points of the radial distorted phase distribution to get the radial phase distribution of the undistorted fringe-pattern. At the same time, the fundamental frequency $f_0$ is determined by the linear fitting parameter $k = 2\pi f_0$ according to Eq. (10).

Step 3: Calculate the radial modulated phase distribution according to Eq. (11), as the blue dash line shown in Fig. 6. Figure 6 shows that the radial modulated phase distribution is almost rotationally symmetric with respect to the central optical axis. Theoretically, we can use the radial modulated phase along the positive direction to calculate the distortion distribution. But in practice, the radial modulated phase calculated is not completely central symmetric. Therefore, the radial modulated phase along the negative direction rotates symmetrically along the center of the modulated phase, then it add to the modulated phase along the positive direction to obtain the average modulated phase along the positive direction, which is to be employed to calculate the distortion distribution.

Step 4: Extend the radial modulated phase distribution to a larger region for image calibration, because the maximum calculating distance for image calibration is half-length of the diagonal line of the distorted image of 1102 pixels. Here, cubic polynomial fitting is employed to extend the radial modulated phase. The numerical fitting parameters from high order items to low order items are:
$-5.8123 \times 10^{-9}, 8.7184 \times 10^{-9}, 7.5508 \times 10^{-8}, -3.9207 \times 10^{-8}$.

Step 5: Calculated the radial distortion value according to Eq. (12). Radial distortion is assumed to be circular symmetric about the center of the image, so the distortion value of each pixel of the distorted image can be calculated. By employing a bilinear interpolation algorithm, image calibration is performed according to Eq. (3) and Eq. (4). Figure 7 shows the calibrated image corresponding to the second distorted image shown in Fig. 3.

We find that it can achieve the image calibration well by the proposed method.

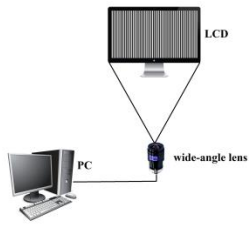

Fig. 2. Experimental apparatus.

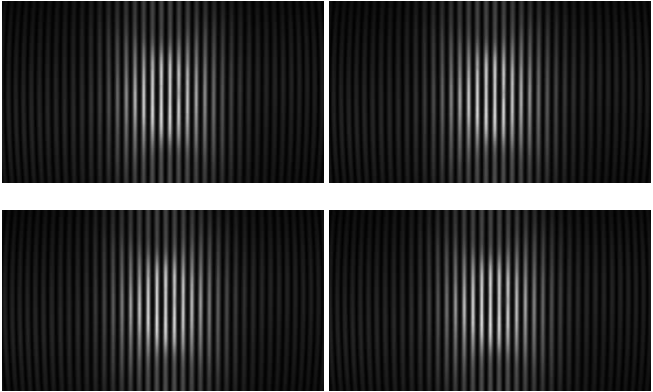

Fig. 3. Four sinusoidal fringes shown on LCD.

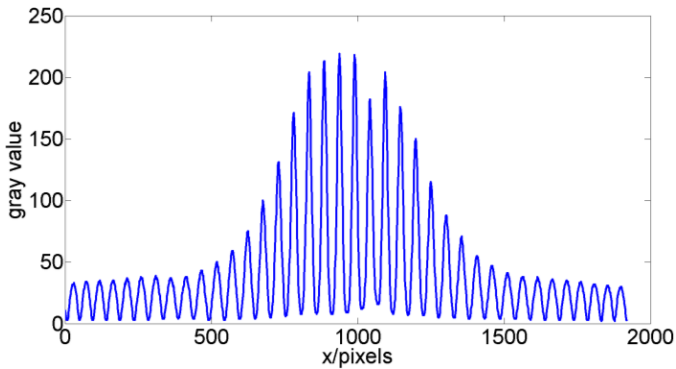

Fig. 4. Gray value distribution.

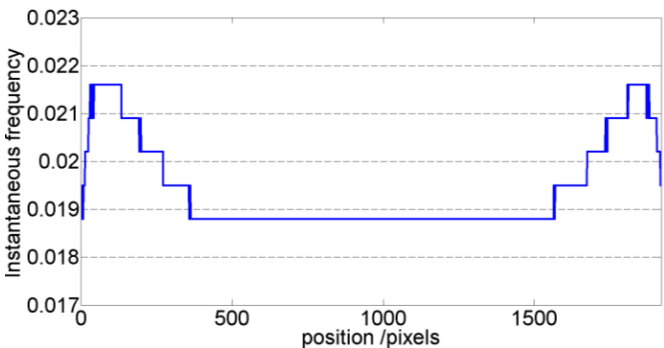

Fig. 5. The instantaneous frequency of the fringe-pattern.

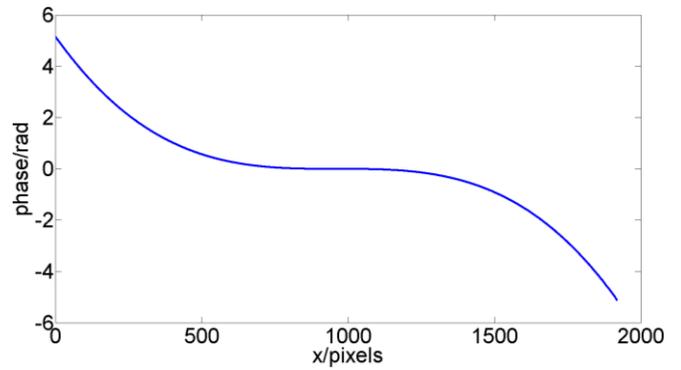

Fig. 6. Experimental radial modulated phase distribution.

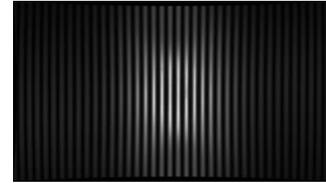

Fig. 7. Experimental calibrated image.

Experiments of different distances from the wide-angle lens to LCD are performed to determine the effect of the recording depth of field on the radial distortion distribution. A series of four sinusoidal fringe-patterns at different positions are captured by the wide-angle lens in turn. The recording distance from the wide-lens to the LCD screen is controlled by the step motor as 25.0cm, 22.0cm, 19.0cm, 16.0cm and 13.0cm. Figure 8 shows the corresponding radial distortion distribution respectively analyzed by the proposed method. We can find that there is little difference from each other. So, we can conclude that the radial distortion of the image is not related to the recording depth of field. It depends on the distance of the pixel point from the origin in the image plane, i.e., it is related to the field of view angle of wide-angle lens imaging system. Larger the distance is, larger the radial distortion value is.

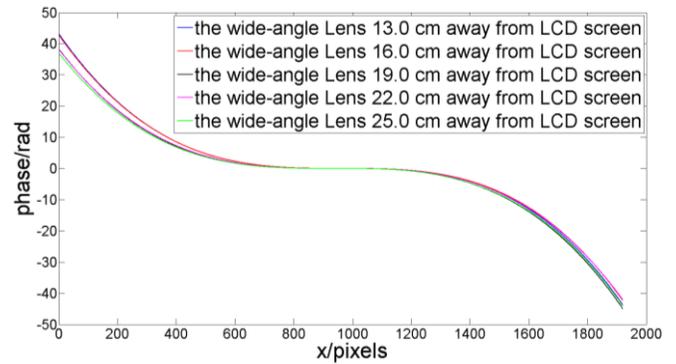

Fig. 8. Radial distortion distribution according to the recording distance of 13.0cm, 16.0cm, 19.0cm, 22.0cm and 25.0cm.

The checkerboard, outdoor scene of close-range and far-range are employed for experiment. Figure 9 shows the distorted checkerboard image captured by the wide-angle lens and the corresponding calibrated image. Figure 10 shows the distorted image of a far-range scene and the corresponding calibrated image. Figure 11 shows the distorted image of a close-range scene and the corresponding calibrated image. The experimental results show that, the distorted images achieve calibration effectively by the proposed method.

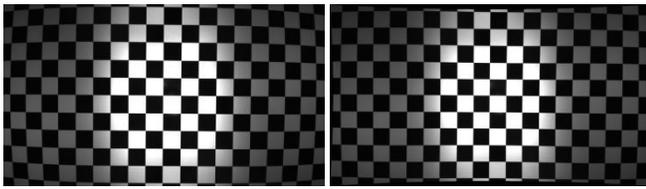
Fig. 9. Distorted checkboard image (a) and the corresponding calibrated image (b).

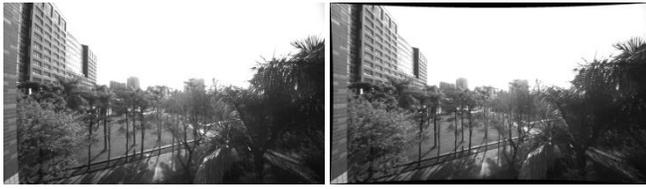
Fig. 10. Distorted image of a far-range scene (a) and the corresponding calibrated image (b).

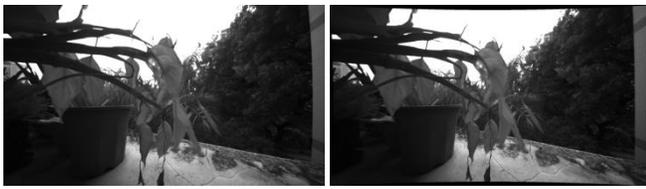
Fig. 11. Distorted image of a close-range scene (a) and the corresponding calibrated image (b).

## 5. CONCLUSION

In this paper, a simple, effective and automatic calibration method for wide-angle lens based on fringe-pattern phase analysis is presented. There are some advantages as following.

Firstly, a simple and effective image distortion calibration is considered according to the experimental result of the radial distortion of the image not related to the recording depth of field, but depending on the field of view angle of the wide-angle lens imaging system. The distortion calibration does not need to establish distortion model and extract feature points or lines either, and can calculate the distortion value of each pixel of the distorted image, correcting the image quickly and accurately.

Secondly, the radial phase distribution of the distorted fringe-pattern is acquired by employing the four-step phase-shifting method for phase de-modulation. The four-step phase-shifting method due to its best spatial localization merit can provide the high precision phase distribution of the distorted fringe-pattern.

Thirdly, the radial phase distribution of the undistorted fringe-pattern is calculated by the numerical linear fitting according to the central 9 points of the radial distorted phase distribution. So, the radial modulated phase is computed. The numerical calculation process is very simple, effective and automatic.

Fourthly, an appropriate and effective fitting model is employed to extend the radial modulated phase distribution to a larger region to avoid the image interception process and achieve calibration for the whole image. By use of the bi-linearity interpolation, we can obtain the calibrated image.

Finally, there is no projecting system in the experimental apparatus. It can avoid projection shadow problems well. And it is simple to perform experiments because of no need to align with the center of the template for distortion measurement.

Theoretical analysis and experimental results show that the proposed method is simple, effective and automatic.

**Funding Information.** National Science Foundation (NSF) (61875074，61971201); Open Fund of the Guangdong Provincial Key Laboratory of Optical Fiber Sensing and Communications (Jinan University)

**Acknowledgements.** We thank Professor Chuping Yang of South China Agricultural University for the help of data processing.